\documentclass[aps,preprint,superscriptaddress]{revtex4-1}

\usepackage{amsmath,amssymb,graphicx}
\usepackage{url}
\usepackage[colorlinks,urlcolor=blue]{hyperref}

\begin{document}
\title{Spin alignment of vector mesons in heavy-ion collisions}
\author{Xin-Li Sheng}
\affiliation{INFN-Firenze, Via Giovanni Sansone, 1, 50019 Sesto Fiorentino FI, Italy}
\affiliation{Peng Huanwu Center for Fundamental Theory and Department of Modern
Physics, University of Science and Technology of China, Hefei, Anhui 230026, China}
\author{Lucia Oliva}
\affiliation{Department of Physics and Astronomy "Ettore Majorana",
University of Catania, Via S. Sofia 64, I-95123 Catania, Italy}
\affiliation{INFN Sezione di Catania, Via S. Sofia 64, I-95123 Catania, Italy}
\author{Zuo-Tang Liang}
\affiliation{Key Laboratory of Particle Physics and Particle Irradiation (MOE),
Institute of Frontier and Interdisciplinary Science, Shandong University,
Qingdao, Shandong 266237, China}
\author{Qun Wang}
\affiliation{Peng Huanwu Center for Fundamental Theory and Department of Modern
Physics, University of Science and Technology of China, Hefei, Anhui 230026, China}
\author{Xin-Nian Wang}
\affiliation{Nuclear Science Division, MS 70R0319, Lawrence Berkeley National Laboratory,
Berkeley, California 94720, USA}
\begin{abstract}
Polarized quarks and antiquarks in high-energy heavy-ion collisions can lead to the spin alignment of vector mesons formed by quark coalescence. Using the relativistic spin Boltzmann equation for vector mesons derived from Kadanoff-Baym equations with an effective quark-meson model for strong interaction and quark coalescence model for hadronizaton, we calculate the spin density matrix element $\rho_{00}$ for $\phi$ mesons and show that anisotropies of local field correlations with respect to the spin quantization direction lead to $\phi$ meson's spin alignment.
We propose that the local correlation or fluctuation of $\phi$ fields is the dominant mechanism for the observed the $\phi$ meson's spin alignment and its strength can be extracted from experimental data
as functions of collision energies. The calculated transverse momentum dependence of $\rho_{00}$ agrees with STAR's data. We further predict the azimuthal angle dependence of $\rho_{00}$ which can be tested in future experiments.

\end{abstract}

\preprint{USTC-ICTS/PCFT-22-16}

\maketitle



{\it Introduction.} 
In noncentral heavy-ion collisions,
the system carries a large  initial orbital angular momentum (OAM) perpendicular to the reaction plane. Part of the OAM can be converted to the vorticity fields of the quark-gluon plasma which in turn lead to the global spin polarization of partons and final hadrons \citep{Liang:2004ph,Liang:2004xn,Voloshin:2004ha,Betz:2007kg,Becattini:2007sr,Gao:2007bc}
(see, e.g. \citep{Wang:2017jpl,Florkowski:2018fap,Becattini:2020ngo,Gao:2020lxh,Huang:2020dtn},
for recent reviews), similar to the Barnett effect \citep{Barnett:1935} and the Einstein-de Haas effect \citep{dehaas:1915}
in materials. The global spin polarization of $\Lambda$ and $\overline{\Lambda}$ hyperons has been observed in Au+Au collisions at $\sqrt{s_{\rm NN}}=$7.7-200 GeV by the STAR Collaboration \citep{STAR:2017ckg,Adam:2018ivw}. According to the quark coalescence model \citep{Liang:2004ph,Yang:2017sdk}, the spin polarization of $\Lambda$ and $\overline{\Lambda}$ is carried by the constituent strange $s$ and antistrange $\overline{s}$ quark, respectively. Therefore, STAR's measurement indicates that $s$ and $\overline{s}$ quarks are also globally polarized along the OAM direction before  hadronization.

Shortly after the prediction of global quark spin polarization \citep{Liang:2004ph} in heavy-ion collisions,  it was also suggested \citep{Liang:2004xn} that the polarized $s$ and $\overline{s}$ quarks can recombine and form polarized vector mesons such as $\phi(1020)$ whose spins align in the OAM direction. For vector mesons, the spin density matrix $\rho_{\lambda_{1}\lambda_{2}}$ is used to describe its spin states with $\lambda_{1},\lambda_{2}=0,\pm1$, labeling the spin state along a specific spin quantization direction. The spin density matrix has unit trace and its diagonal elements are probabilities for spin states with $\lambda=0,\pm 1$. However, the spin polarization of vector mesons, proportional to $\rho_{11}-\rho_{-1,-1}$, cannot be directly measured through strong interaction decays. Instead, $\rho_{00}$ can be measured
through the angular distribution of its strong decay daughters \citep{Liang:2004xn,Yang:2017sdk,Tang:2018qtu,Goncalves:2021ziy,Mohanty:2021vbt}. On average, the polarization vector $\epsilon ^\mu (\lambda)$ of vector mesons is in the plane perpendicular to the spin quantization direction if $\rho_{00}<$1/3, while it is aligned in the quantization direction if $\rho_{00}>$1/3.

Such spin alignment of the $\phi$ meson in the OAM direction was indeed observed recently by STAR experiment \citep{STAR:2022fan}. However, the measured positive deviation from 1/3 of $\rho_{00}^\phi$ is orders of magnitude larger than what one would expect from the same vorticity that causes the measured $\Lambda$ and $\overline{\Lambda}$ polarization in the same collisions. Contributions from electromagnetic fields and other possible conventional mechanisms are also orders of magnitude smaller \citep{Yang:2017sdk,Sheng:2019kmk,Xia:2020tyd,Gao:2021rom,Muller:2021hpe}.

In this Letter we propose that the local fluctuations or correlations of the $\phi$ meson fields during hadronization can be responsible for the observed spin alignment of the final-state $\phi$ meson  in the framework of relativistic quantum transport theory. The effect of $\phi$'s mean field on the hyperon polarization was proposed before \citep{Csernai:2018yok}. But its magnitude should be very small implied by the negligible difference between the observed global polarization of $\Lambda$ and $\overline{\Lambda}$ \citep{STAR:2017ckg,Adam:2018ivw},
since the sign of the polarization by the $\phi$ field is opposite for $s$ and $\overline{s}$ quarks \citep{Sheng:2019kmk}.
Using the relativistic spin Boltzmann equation for vector mesons derived in this study, we will show that the deviation from 1/3 of the spin density matrix $\rho_{00}$ is proportional to the spatially anisotropic short distance correlations or fluctuations of the vector meson fields. One can therefore extract the strength of the field fluctuations or correlations from the experimental data and predict the transverse momentum and azimuthal angle dependence of the spin alignment.




{\it Spin Boltzmann equation and spin density matrices.}
Nonrelativistic quark coalescence or recombination models have been employed
to describe $\rho_{00}$ from the spin polarization of the quark and antiquark \citep{Liang:2004xn,Yang:2017sdk,Sheng:2019kmk}.
Recently the spin Boltzmann equation (SBE) for vector mesons has been derived by us  in the framework of relativistic quantum transport theory \citep{Sheng:2022ffb}.  At the leading order in $\hbar$, the collision terms can be expressed in terms of matrix-valued spin dependent distributions (MVSDs) of the quark, antiquark \citep{Becattini:2013fla}, and vector meson \citep{Sheng:2021kfc} in the effective quark-meson model \citep{Manohar:1983md,Fernandez:1993hx,Li:1997gd,Zhao:1998fn,Zacchi:2015lwa,Zacchi:2016tjw} for strong interaction during hadronization. This provides a more rigorous framework to calculate spin observables for vector mesons  such as $\rho_{00}$ for the $\phi$ meson.

The Wigner functions for massless vector particles such as gluons and photons \citep{Elze:1986hq,Blaizot:2001nr,Wang:2001dm,Huang:2020kik,Hattori:2020gqh,Muller:2021hpe} have been studies for many years, but to our knowledge there are few works about Wigner functions for massive vector mesons in the context of spin polarization (see Ref. \citep{Weickgenannt:2022jes} for a recent study). From the Kadanoff-Baym equation \citep{Kadanoff:1962,Blaizot:2001nr,Berges:2004yj,Cassing:2008nn}
for Wigner functions, the spin Boltzmann equation for the vector meson's
MVSD $f_{\lambda_{1}\lambda_{2}}^{V}$ with coalescence and dissociation collision terms reads \citep{Sheng:2022ffb},
\begin{align}
k\cdot\partial_{x}f_{\lambda_{1}\lambda_{2}}^{V}(x,\mathbf{k}) & =\frac{1}{16}\sum_{\lambda_{1}^{\prime},\lambda_{2}^{\prime}}\left[\epsilon_{\mu}^{\ast}(\lambda_{1},{\bf k})\epsilon_{\nu}(\lambda_{1}^{\prime},{\bf k})\delta_{\lambda_{2}\lambda_{2}^{\prime}}\right.\nonumber \\
 & \hspace{-0.5cm}\left.+\delta_{\lambda_{1}\lambda_{1}^{\prime}}\epsilon_{\mu}^{\ast}(\lambda_{2}^{\prime},{\bf k})\epsilon_{\nu}(\lambda_{2},{\bf k})\right]\mathcal{C}_{\lambda_{1}^{\prime}\lambda_{2}^{\prime}}^{\mu\nu}(x,{\bf k}),\label{eq:boltz-mvsd-1}
\end{align}
where $\lambda_{1}$, $\lambda_{2}$, $\lambda_{1}^{\prime}$, and
$\lambda_{2}^{\prime}$ denote the spin states of vector mesons along
the spin quantization direction.
We consider coalescence as the main process for primary
particle production in heavy-ion collisions \citep{Greco:2003xt,Fries:2003vb,Greco:2003mm,Fries:2003kq,Greco:2003vf,Zhao:2020wcd}. The collision kernel $\mathcal{C}_{\lambda_{1}^{\prime}\lambda_{2}^{\prime}}^{\mu\nu}(x,{\bf k})$ is an integral over the quark's and antiquark's momenta which contains in the integrand a delta function for energy-momentum conservation, a gain and a loss term involving MVSDs for the quark, antiquark and vector meson, $f_{rs}^{q}$, $f_{rs}^{\overline{q}}$ and $f_{\lambda_{1}^{\prime}\lambda_{2}^{\prime}}^{V}$ respectively,
and a matrix element squared involving Dirac spinors of the quark and antiquark with spin indices. One can find the explicit form of $\mathcal{C}_{\lambda_{1}^{\prime}\lambda_{2}^{\prime}}^{\mu\nu}(x,{\bf k})$ in Ref. \citep{Sheng:2022ffb}. In the matrix element squared there are also $q\overline{q}V$ vertices in the form $\Gamma^{\alpha}\approx g_{V}B(\mathbf{k}-\mathbf{p}^{\prime},\mathbf{p}^{\prime})\gamma^{\alpha}$,
where $g_{V}$ is the coupling constant of the vector meson and quark-antiquark, and $B(\mathbf{k}-\mathbf{p}^{\prime},\mathbf{p}^{\prime})$ denotes the covariant Bethe-Salpeter wave function of the vector meson \citep{Xu:2019ilh,Xu:2021mju}. Note that $f_{rs}^{q}$ and $f_{rs}^{\overline{q}}$ ($r$ and $s$ denote spin indices) are related to the spin polarization four-vectors of quark and anti-quark \citep{Becattini:2013fla,Weickgenannt:2020aaf,Sheng:2021kfc,Weickgenannt:2021cuo}, $P_{q}^{\mu}$ and $P_{\overline{q}}^{\mu}$, respectively,
\begin{equation}
f_{rs}^{q(\overline{q})}(x,\mathbf{p})=\frac{1}{2}f_{q(\overline{q})}(x,\mathbf{p})\left[\delta_{rs}-P_{\mu}^{q(\overline{q})}(x,\mathbf{p})n_{j}^{\mu}(\mathbf{p})\tau_{rs}^{j}\right],\label{eq:f-rs-pol}
\end{equation}
where $f_{q(\overline{q})}(x,\mathbf{p})$ is the unpolarized distribution for the
quark (antiquark), $n_{j}^{\mu}(\mathbf{p})$ ($j=1,2,3$) are four-vectors of
three basis directions for spin states in the (anti-)quark's rest frame with
the $j=3$ component denoting the spin quantization direction \citep{Sheng:2021kfc}, and
$\tau^{j}$ ($j=1,2,3$) denote three Pauli matrices in the space
of spin states denoted by $r$ and $s$.

The gain and loss terms in $\mathcal{C}_{\lambda_{1}^{\prime}\lambda_{2}^{\prime}}^{\mu\nu}(x,{\bf k})$ correspond
to the coalescence and dissociation processes, respectively. During the hadronization
stage of heavy-ion collisions, the distribution functions for vector mesons and constituent (anti-)quarks are normally much less than 1, which allows us to take the dilute gas limit $f_{\lambda_{1}\lambda_{2}}^{V}\sim f_{rs}^{q}\sim f_{rs}^{\overline{q}}\ll1$.
Then Eq. (\ref{eq:boltz-mvsd-1}) can be expressed as
\begin{align}
k\cdot\partial_{x}f_{\lambda_{1}\lambda_{2}}^{V}(x,\mathbf{k})= & \frac{1}{8}\left[\epsilon_{\mu}^{\ast}(\lambda_{1},{\bf k})\epsilon_{\nu}(\lambda_{2},{\bf k})\mathcal{C}_{\text{coal}}^{\mu\nu}(x,{\bf k})\right.\nonumber \\
 & \hspace{0.5cm}\left.-\mathcal{C}_{\text{diss}}({\bf k})f_{\lambda_{1}\lambda_{2}}^{V}(x,\mathbf{k})\right],\label{eq:Bltzmann}
\end{align}
where the dissociation kernel $\mathcal{C}_{\text{diss}}$ is independent
of the MVSDs. The coalescence kernel $\mathcal{C}_{\text{coal}}^{\mu\nu}$
can be obtained by substituting the MVSDs for quarks and antiquarks
into the gain term of $\mathcal{C}_{\lambda_{1}^{\prime}\lambda_{2}^{\prime}}^{\mu\nu}(x,{\bf k})$ and carrying out
a summation over spin indices of the quark and antiquark,
\begin{align}
\mathcal{C}_{\text{coal}}^{\mu\nu}(x,{\bf k}) & =\int\frac{d^{3}\mathbf{p}^{\prime}}{(2\pi\hbar)^{2}}\frac{1}{E_{{\bf p}^{\prime}}^{\overline{q}}E_{{\bf k}-{\bf p}^{\prime}}^{q}}\delta\left(E_{{\bf k}}^{V}-E_{{\bf p}^{\prime}}^{\overline{q}}-E_{{\bf k}-{\bf p}^{\prime}}^{q}\right)\nonumber \\
 & \hspace{-0.5cm}\times\text{Tr}\left\{ \Gamma^{\nu}\left(p^{\prime}\cdot\gamma-m_{\overline{q}}\right)\left[1+\gamma_{5}\gamma\cdot P^{\overline{q}}(x,\mathbf{p}^{\prime})\right]\right.\nonumber \\
 & \hspace{-0.5cm}\left.\times\Gamma^{\mu}\left[(k-p^{\prime})\cdot\gamma+m_{q}\right]\left[1+\gamma_{5}\gamma\cdot P^{q}(x,\mathbf{k}-\mathbf{p}^{\prime})\right]\right\} \nonumber \\
 & \hspace{-0.5cm}\times f_{\overline{q}}(x,\mathbf{p}^{\prime})f_{q}(x,{\bf k}-\mathbf{p}^{\prime}),\label{eq:coal_term}
\end{align}
where $k^{\mu}=(E_{{\bf k}}^{V},{\bf k})$ and $p^{\prime\mu}=(E_{{\bf p}^{\prime}}^{\overline{q}},{\bf p}^{\prime})$
denote the on-shell four-momenta of the vector meson and the antiquark respectively,
and $m_{q}=m_{\overline{q}}$ are masses for the quark and antiquark.


Schematically, the formal solution to Eq. (\ref{eq:Bltzmann}) reads
\begin{align}
f_{\lambda_{1}\lambda_{2}}^{V}(x,\mathbf{k})\sim & \frac{1}{\mathcal{C}_{\text{diss}}({\bf k})}\left[1-e^{-\mathcal{C}_{\text{diss}}({\bf k})\Delta t}\right]\nonumber \\
 & \times\epsilon_{\mu}^{\ast}(\lambda_{1},{\bf k})\epsilon_{\nu}(\lambda_{2},{\bf k})\mathcal{C}_{\text{coal}}^{\mu\nu}(x,{\bf k}),
\end{align}
where $\Delta t$ is the formation time of the vector meson and we
assume $f_{\lambda_{1}\lambda_{2}}^{V}(x,\mathbf{k})$ is zero at
the initial time. We note that $f_{\lambda_{1}\lambda_{2}}^{V}(x,\mathbf{k})$ is actually
the unnormalized spin density matrix, from which the normalized one, $\rho_{\lambda_{1}\lambda_{2}}$,
is given as
\begin{equation}
\rho_{\lambda_{1}\lambda_{2}}(x,\mathbf{k})=\frac{\epsilon_{\mu}^{\ast}(\lambda_{1},{\bf k})\epsilon_{\nu}(\lambda_{2},{\bf k})\mathcal{C}_{\text{coal}}^{\mu\nu}(x,{\bf k})}{\sum_{\lambda=0,\pm1}\epsilon_{\mu}^{\ast}(\lambda,{\bf k})\epsilon_{\nu}(\lambda,{\bf k})\mathcal{C}_{\text{coal}}^{\mu\nu}(x,{\bf k})}. \label{eq:vector_meson_rho}
\end{equation}
We see that $\rho_{\lambda_{1}\lambda_{2}}$
is fully determined by the coalescence kernel $\mathcal{C}_{\text{coal}}^{\mu\nu}(x,{\bf k})$.
The vector meson's spin density matrix depends on the spin states of its constituent quark and antiquark, similarly as in nonrelativistic coalescence models \citep{Yang:2017sdk,Sheng:2019kmk}.


{\it Spin alignment for $\phi$ mesons.}
In order to apply Eq. (\ref{eq:vector_meson_rho}) to calculate $\rho_{00}$ for the $\phi$ meson from coalescence of $s$ and $\overline{s}$ quarks,
we assume that the chemical freeze-out occurs shortly after the $\phi$ meson's formation through coalescence, so one can neglect the effect from hadronic interaction on its spin states.
We also neglect polarization mechanisms such as by electromagnetic fields or fluid gradients \citep{Wu:2019eyi,Liu:2021uhn,Becattini:2021suc,Fu:2021pok,Becattini:2021iol,Yi:2021ryh} which are not essential in our study here. We will only consider spin polarization by the vorticity field $\omega^{\mu\nu}$ and the $\phi$ field $F_{\rho\sigma}^{\phi}$. Effects from vorticity fields are also negligible \citep{Sheng:2019kmk} but still included just as a contrast in the formalism to $\phi$ fields.
The spin polarization four-vectors as phase space distributions for $s$ and $\overline{s}$ are given by \citep{Becattini:2013fla,Becattini:2016gvu,Fang:2016vpj,Yang:2017sdk,Weickgenannt:2019dks},
\begin{align}
P_{s}^{\mu}(x,{\bf p})\approx & \frac{1}{4m_{s}}\epsilon^{\mu\nu\rho\sigma}\left(\omega_{\rho\sigma}+\frac{g_{\phi}}{(u\cdot p)T_{\mathrm{h}}}F_{\rho\sigma}^{\phi}\right)p_{\nu},\nonumber \\
P_{\overline{s}}^{\mu}(x,{\bf p}) \approx & \frac{1}{4m_{s}}\epsilon^{\mu\nu\rho\sigma}\left(\omega_{\rho\sigma}-\frac{g_{\phi}}{(u\cdot p)T_{\mathrm{h}}}F_{\rho\sigma}^{\phi}\right)p_{\nu},\label{eq:polarization-ssbar}
\end{align}
where $p^{\mu}=(E_{p},{\bf p})$ denotes the on-shell four-momentum of $s$ or $\overline{s}$, $g_\phi$ is the effective coupling constant of the $s\overline{s}\phi$ vertex, $T_{\mathrm{h}}$ is the local temperature when $\phi$ mesons are formed through coalescence, and $f_s$ and $f_{\overline{s}}$ are neglected as compared to 1 at $T_{\rm h}$.
In Eq. (\ref{eq:polarization-ssbar}) we have introduced a reference frame vector $u^{\mu}$, which ensures $P_{s/\overline{s}}^{\mu}$ to be Lorentz pseudovectors. Usually $u^{\mu}$ is taken as the local fluid velocity. In the calculation, we will take $u^{\mu}=(1,0,0,0)$ in the rest frame of the $\phi$ meson for simplicity.

Substituting Eq.~(\ref{eq:polarization-ssbar}) into Eqs.~(\ref{eq:coal_term}) and (\ref{eq:vector_meson_rho}) and by a lengthy but straightforward calculation~\citep{Sheng:2022ffb}, we obtain $\rho_{00}$
in $\phi$ meson's rest frame,
\begin{align}
\rho_{00}(x,{\bf k})\approx & \frac{1}{3}+C_{1}\left[\frac{1}{3}\boldsymbol{\omega}^{\prime}\cdot\boldsymbol{\omega}^{\prime}-(\boldsymbol{\epsilon}_{0}\cdot\boldsymbol{\omega}^{\prime})^{2}\right]\nonumber \\
 & +C_{2}\left[\frac{1}{3}\boldsymbol{\varepsilon}^{\prime}\cdot\boldsymbol{\varepsilon}^{\prime}-(\boldsymbol{\epsilon}_{0}\cdot\boldsymbol{\varepsilon}^{\prime})^{2}\right]\nonumber \\
 & -\frac{4g_{\phi}^{2}}{m_{\phi}^{2}T_{\mathrm{h}}^{2}}C_{1}\left[\frac{1}{3}{\bf B}_{\phi}^{\prime}\cdot{\bf B}_{\phi}^{\prime}-(\boldsymbol{\epsilon}_{0}\cdot{\bf B}_{\phi}^{\prime})^{2}\right]\nonumber \\
 & -\frac{4g_{\phi}^{2}}{m_{\phi}^{2}T_{\mathrm{h}}^{2}}C_{2}\left[\frac{1}{3}{\bf E}_{\phi}^{\prime}\cdot{\bf E}_{\phi}^{\prime}-(\boldsymbol{\epsilon}_{0}\cdot{\bf E}_{\phi}^{\prime})^{2}\right], \label{eq:rho00-cms}
\end{align}
where $\boldsymbol{\epsilon}_{0}$ denotes the spin quantization direction for the $\phi$ meson, $\boldsymbol{\varepsilon}^\prime$ and $\boldsymbol{\omega}^\prime$ denote the electric and magnetic part of $\omega_{\mu\nu}^\prime$,
${\bf E}_{\phi}^\prime$ and ${\bf B}_{\phi}^\prime$ the electric and magnetic part of $F_{\phi}^{\prime\mu\nu}$ in the meson's rest frame.
$C_{1}$ and $C_{2}$ are two coefficients depending only on quark and meson masses,
\begin{align}
C_{1}= & \frac{8m_{s}^{4}+16m_{s}^{2}m_{\phi}^{2}+3m_{\phi}^{4}}{120m_{s}^{2}(m_{\phi}^{2}+2m_{s}^{2})},\nonumber \\
C_{2}= & \frac{8m_{s}^{4}-14m_{s}^{2}m_{\phi}^{2}+3m_{\phi}^{4}}{120m_{s}^{2}(m_{\phi}^{2}+2m_{s}^{2})}. \label{eq:c1c2}
\end{align}
The above simple and highly nontrivial result is remarkable in that all mixed terms of different components of the vorticity and $\phi$ fields disappear due to exact cancellation for quarkonium vector mesons. What remain are short-distance correlations between same components of vorticity and meson fields.
If we neglect variations of these fields within the hadron size, these field correlations become local field fluctuations during the hadronization. Any spatial anisotropy of these field fluctuations in the meson's rest frame with respect to the spin quantization direction $\boldsymbol{\epsilon}_{0}$  will lead to the spin alignment, i.e. $\rho_{00}\neq 1/3$.

In order to calculate the momentum dependence of $\rho_{00}$, one can express it in terms of fields in the lab frame through the Lorentz transformation of $\omega^{\mu\nu}$ and $F_{\phi}^{\mu\nu}$ with the boost factor $\gamma=E_{{\bf k}}^{\phi}/m_{\phi}$ and the $\phi$ meson's velocity ${\bf v}=\mathbf{k}/E_{{\bf k}}^{\phi}$. As a result, $\rho_{00}$ can be expressed by a factorization form $\rho_{00}(x,{\bf k})=1/3+\sum _{i,\alpha} I_i^\alpha({\bf k})O_i^\alpha(x)$, where $I_i^\alpha({\bf k})$ ($i=x,y,z$) denote momentum-dependent functions and $O_i^\alpha(x)$ field fluctuations, $O_i^{\alpha=1-4}(x)=\boldsymbol{\varepsilon}_{i}^{2}$, $\boldsymbol{\omega}_{i}^{2}$, $(g_\phi\mathbf{E}_{i}^{\phi}/T_{\mathrm{h}})^2$ and $(g_\phi\mathbf{B}_{i}^{\phi}/T_{\mathrm{h}})^2$, respectively. To obtain the observed $\rho_{00}$ in experiments, we have to take the space-time average of $O_i^\alpha(x)$ on the hadronization hyper-surface 
and the momentum average of $I_i^\alpha({\bf k})$ weighted by $\phi$ meson's momentum spectra including the azimuthal anisotropy through the elliptic flow $v_2(\bf k)$ given by experimental data.


{\it Extracting field fluctuations and predictions.}
Since one can safely neglect contributions from local vorticities to the $\phi$ meson's spin alignment as compared to the experimental data \citep{Sheng:2019kmk}, the dominant contributions can come from the $\phi$ field's fluctuations in terms of six parameters $\langle (g_\phi\mathbf{B}_{i}^{\phi}/T_{\mathrm{h}})^{2}\rangle$
and $\langle (g_\phi\mathbf{E}_{i}^{\phi}/T_{\mathrm{h}})^{2}\rangle$ ($i=x,y,z$). Considering the geometry of the fireball in heavy-ion collisions,  we assume that the fluctuations of transverse and longitudinal fields are different, as represented by
$\left\langle (g_\phi\mathbf{B}_{x,y}^{\phi}/T_{\mathrm{h}})^{2}\right\rangle
=\left\langle (g_\phi\mathbf{E}_{x,y}^{\phi}/T_{\mathrm{h}})^{2}\right\rangle \equiv F_T^{2}$ and
$\left\langle (g_\phi\mathbf{B}_{z}^{\phi}/T_{\mathrm{h}})^{2}\right\rangle
=\left\langle (g_\phi\mathbf{E}_{z}^{\phi}/T_{\mathrm{h}})^{2}\right\rangle \equiv F_z^{2}$.
Such an assumption is consistent with the numerical estimates of the usual electromagnetic  fields \citep{Voronyuk:2011jd,Deng:2012pc,Li:2016tel,Siddique:2021smf}.

\begin{figure}
\includegraphics[width=8.5cm]{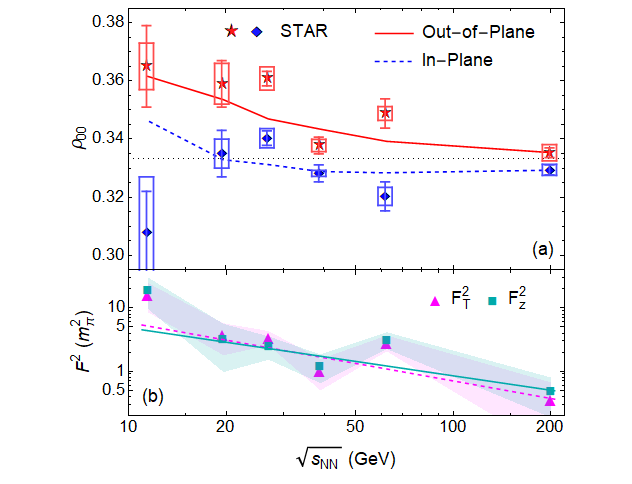}
\caption{\label{fig:rho00-eng}(a) The STAR's data \cite{STAR:2022fan} on $\phi$ meson's $\rho_{00}^y$ (out-of-plane, red stars) and $\rho_{00}^x$ (in-plane, blue diamonds) in 0-80\% Au+Au collisions as functions of collision energies. The red-solid line (out-of-plane) and blue-dashed line (in-plane) are calculated with values of $F_T^{2}$ and $F_z^{2}$ from fitted curves in (b). (b) Values of $F_T^{2}$ (magenta triangles) and $F_z^{2}$ (cyan squares) with shaded error bands extracted from the STAR's data on the $\phi$ meson's $\rho_{00}^y$ and $\rho_{00}^x$ in (a). The magenta-dashed line  (cyan-solid line) is a fit to the extracted  $F_{T}^{2}$ ($F_{z}^{2}$) as a function of $\sqrt{s_{\mathrm{NN}}}$ (see the text).
}
\end{figure}

We can determine the two parameters on field fluctuations by fitting the STAR data on momentum-integrated
$\rho_{00}^y$ (out of plane) and $\rho_{00}^{x}$ (in plane),
corresponding to the spin quantization direction $\boldsymbol{\epsilon}_0=(0,1,0)$ and (1,0,0), respectively.  In our calculation of the momentum-integrated $\rho_{00}^y$ and $\rho_{00}^{x}$, we have used
the $\phi$ meson's transverse momentum spectra and $v_2(k_T)$ from STAR's experiments at $\sqrt{s_{\rm NN}}=11.5$-$200$ GeV \citep{STAR:2008bgi,STAR:2019bjj,STAR:2007mum,STAR:2013ayu,STAR:2008bgi}
as the weight function in ranges of $k_{T}=$1.2-5.4 GeV and rapidity $|y|<1$.  The difference between $\rho_{00}^y$ and $\rho_{00}^{x}$ is driven by the momentum anisotropy via $v_2(k_T)$.  We will consider 0-80\% Au+Au collisions in all our calculations and comparisons with the experimental data in this study. Since there are no data available for $\phi$ meson's $k_T$ spectra in 0-80\% Au+Au collisions, we will use the data in 30-40\% centrality instead. Since the weighted momentum average is only sensitive to the shape of the spectra, the errors from such substitute should be small. The constituent quark mass is set to $m_s=419$ MeV \citep{Godfrey:1985xj} with $m_\phi=1020$ MeV.
The fits to the STAR's data on the momentum-integrated $\rho_{00}^y$ and $\rho_{00}^{x}$ and the extracted values of $F_T^{2}$ and $F_z^{2}$ as functions of colliding energies are shown in Fig.~\ref{fig:rho00-eng}.
The energy dependence of $F_T^{2}$ and $F_z^{2}$ can be fitted with a function $\ln (F_{T,z}^2/m_\pi^2)=a_{T,z} - b_{T,z} \ln (\sqrt{s_{\rm NN}}/\mathrm{GeV})$, with $a_T=3.90\pm 1.11$, $b_T=0.924\pm 0.234$, $a_z=3.33\pm 0.917$, and $b_z=0.760\pm 0.189$. The shaded areas in Fig.~\ref{fig:rho00-eng}(b) reflect errors of the momentum-integrated $\rho_{00}^y$ and $\rho_{00}^{x}$ in STAR's measurement. Errors of the STAR data for $\phi$'s spectra and $v_2$ are negligible in extracting $F_T^{2}$ and $F_z^{2}$ as compared to those of $\rho_{00}$ and will be omitted in the following calculations.
A variation of $m_s$ from 419 to 486 MeV \citep{Griffiths:2008zz} gives an increase in the extracted values of $F_T^2$ and $F_z^2$ by about 37\% through $C_1$ and $C_2$ in Eq. (\ref{eq:c1c2}).


\begin{figure}
\includegraphics[width=7cm]{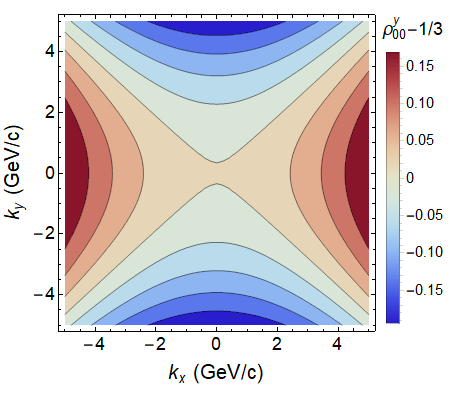}
\caption{\label{fig:rho00-px-py}
Contour plot of $\rho_{00}^{y}-1/3$ for $\phi$ mesons as a function of $k_{x}$ and $k_{y}$ in 0-80\% Au+Au collisions at $\sqrt{s_{\mathrm{NN}}}=$200 GeV.}
\end{figure}

\begin{figure}
\includegraphics[width=7cm]{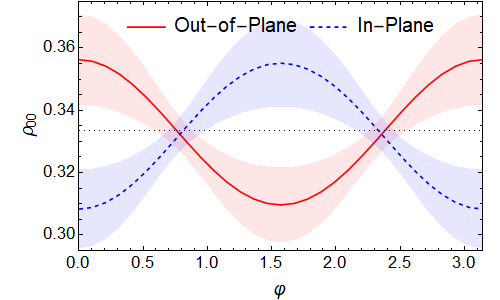}
\caption{\label{fig:rho00-phi}
Calculated $\rho_{00}^{y}$ (out-of-plane) and $\rho_{00}^{x}$ (in-plane)
of $\phi$ mesons as functions of the azimuthal angle $\varphi$ in 0-80\% Au+Au collisions at $\sqrt{s_{\mathrm{NN}}}$=200 GeV. Shaded error bands are from the extracted parameters $F_T^2$ and $F_z^2$. }
\end{figure}

With the extracted values of $F_T^{2}$ and $F_z^{2}$ at each colliding energy, we can look at the transverse momentum and azimuthal angle dependence of $\rho_{00}(\mathbf{k})$. In Fig. \ref{fig:rho00-px-py}, we show  the contour plot of $\rho_{00}^{y}-1/3$ in $k_{x}$ and $k_{y}$ at $\sqrt{s_{\rm NN}}=200$ GeV, averaged over the central rapidity region $|y|<1$.
We can see a strong modulation of $\rho_{00}^{y}$ in the azimuthal angle.
If we integrate $\rho_{00}(\mathbf{k})$ over $k_T$ weighted by its spectra in the range $k_{T}=$1.2-5.4 GeV, we can obtain the modulation of $\rho_{00}^{y}$ and $\rho_{00}^{x}$ with the azimuthal angle $\varphi$ in Fig.~\ref{fig:rho00-phi}. This is an interesting model prediction for future experimental verification.

\begin{figure}
\includegraphics[width=7.5cm]{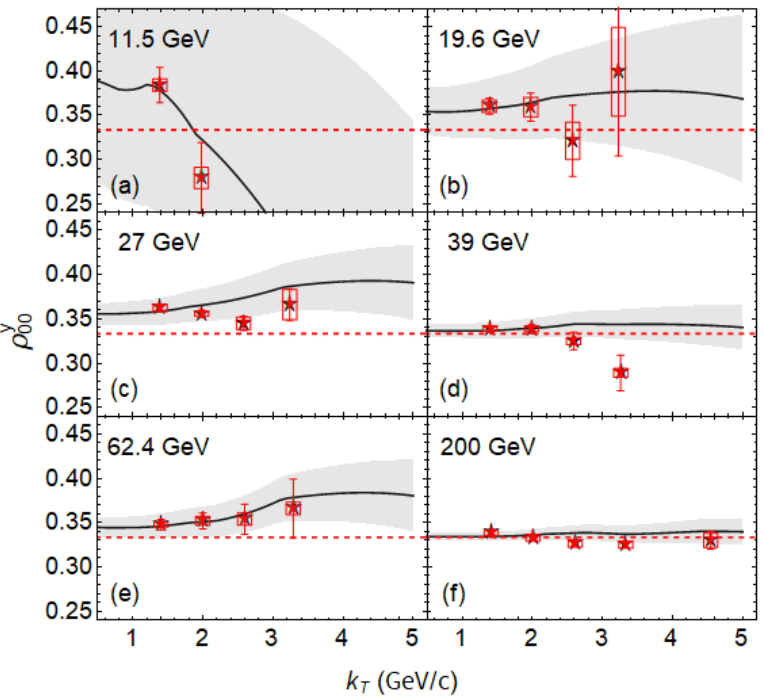}
\caption{\label{fig:rho00-pt} Calculated $\rho_{00}^{y}$ for $\phi$ mesons (solid lines) as functions of transverse momenta in 0-80\% Au+Au collisions at different colliding energies as compared to STAR data \cite{STAR:2022fan}. Shaded error bands are from the extracted parameters $F_T^2$ and $F_z^2$.}
\end{figure}

Averaging over the azimuthal angle at fixed $k_T$ and using the $v_2(k_T)$ data to describe the azimuthal anisotropy, we obtain the $k_{T}$ dependence of $\rho_{00}^{y}$ in Fig. \ref{fig:rho00-pt} as compared to STAR's data for six colliding energies (11.5, 19.6, 27, 39, 62.4, 200 GeV). For large $k_T$ beyond the range of the $v_2(k_T)$ data, we use a linear extrapolation between the data value of $v_2$ at the highest $k_T$ and $v_2=$0 at a larger $k_T$ outside the experimental range which we set to 10 GeV/$c$. The error bands in the calculation are mainly due to those of the two parameters $F_T^2$ and $F_z^2$ extracted from experimental data at each colliding energy. We find that our predicted $\rho_{00}^{y}$ is nearly a constant at $k_{T}<2$ GeV and increases slightly at higher $k_T$.


{\it Summary.}
Based on a relativistic quantum transport theory for spin dynamics,
we have formulated the spin density matrix element $\rho_{00}$ for $\phi$ mesons employing the spin Boltzmann equation with the effective quark-meson model for interaction and quark coalescence model for hadronization. Neglecting effects of hadronic interaction after the hadronization, the final $\rho_{00}-1/3$ is found to be proportional to local correlations or fluctuations of the $\phi$ field. The effective $\phi$ field's fluctuation parameters can be extracted through comparison with the STAR data on momentum-integrated $\rho_{00}$. Their values and colliding energy dependence may shed light on non-perturbative properties of strong interaction \citep{Shuryak:1981ff,Shuryak:1982dp,Diakonov:1985eg,Shifman:1978bx,Shifman:1978by,Schafer:1996wv,Kharzeev:1999vh}. We further predicted the transverse momentum and azimuthal angle dependence of $\rho_{00}$ that can be verified by future experiments.
Our theoretical method can also be applied to the spin alignment of heavy quarkonia \citep{ALICE:2022sli} and spin correlation of hyperons \citep{Gong:2021bcp}.

\section*{Acknowledgement}

The authors thank C.D. Roberts for providing us with the Bethe-Salpeter
wave function of the $\phi$ meson. The authors thank X.G. Huang, J.F. Liao, S. Pu, A.H. Tang, Di-Lun Yang and Y. Yin for helpful discussion. This work was supported in part by the NSFC under Grants Nos. 12135011, 11890713, 11890710 and 11890714, by the Strategic Priority Research Program of the Chinese Academy of Sciences (CAS) under Grant No. XDB34030102 and by U.S. DOE under Contract No. DE-AC02-05CH11231.

\bibliographystyle{apsrev}
\bibliography{ref-kb-vm-short}

\end{document}